\documentclass[superscriptaddress,aps,preprintnumbers,amsmath,showpacs,amssymb,prd,nofootinbib,reprint]{revtex4-1}
\usepackage{bm, color} 
\usepackage{amssymb,amsfonts,slashed,amsthm,amsmath,graphicx, soul}
\usepackage[caption=false]{subfig}

\begin{document}


\renewcommand{\figurename}{Fig.}
\renewcommand{\tablename}{Table.}
\newcommand{\Slash}[1]{{\ooalign{\hfil#1\hfil\crcr\raise.167ex\hbox{/}}}}
\newcommand{\bra}[1]{ \langle {#1} | }
\newcommand{\ket}[1]{ | {#1} \rangle }
\newcommand{\bef}{\begin{figure}}  \newcommand{\eef}{\end{figure}}
\newcommand{\bec}{\begin{center}}  \newcommand{\eec}{\end{center}}
\newcommand{\laq}[1]{\label{eq:#1}}  
\newcommand{\dd}[1]{{d \o d{#1}}}
\newcommand{\Eq}[1]{Eq.~(\ref{eq:#1})}
\newcommand{\Eqs}[1]{Eqs.~(\ref{eq:#1})}
\newcommand{\eq}[1]{(\ref{eq:#1})}
\newcommand{\Sec}[1]{Sec.\ref{chap:#1}}
\newcommand{\ab}[1]{\left|{#1}\right|}
\newcommand{\vev}[1]{ \left\langle {#1} \right\rangle }
\newcommand{\bs}[1]{ {\boldsymbol {#1}} }
\newcommand{\lac}[1]{\label{chap:#1}}
\newcommand{\SU}[1]{{\rm SU{#1} } }
\newcommand{\SO}[1]{{\rm SO{#1}} }
\def\({\left(}
\def\){\right)}
\def\dt{{d \o dt}}
\def\diag{\mathop{\rm diag}\nolimits}
\def\Spin{\mathop{\rm Spin}}
\def\O{\mathcal{O}}
\def\U{\mathop{\rm U}}
\def\Sp{\mathop{\rm Sp}}
\def\SL{\mathop{\rm SL}}
\def\tr{\mathop{\rm tr}}
\def\ebq{\end{equation} \begin{equation}}
\newcommand{\OR}{~{\rm or}~}
\newcommand{\AND}{~{\rm and}~}
\newcommand{\EV}{ {\rm \, eV} }
\newcommand{\KEV}{ {\rm \, keV} }
\newcommand{\MEV}{ {\rm \, MeV} }
\newcommand{\GEV}{ {\rm \, GeV} }
\newcommand{\TEV}{ {\rm \, TeV} }
\def\o{\over}
\def\a{\alpha}
\def\b{\beta}
\def\c{\varepsilon}
\def\d{\delta}
\def\e{\epsilon}
\def\f{\phi}
\def\g{\gamma}
\def\h{\theta}
\def\k{\kappa}
\def\l{\lambda}
\def\m{\mu}
\def\n{\nu}
\def\p{\psi}
\def\q{\partial}
\def\r{\rho}
\def\s{\sigma}
\def\t{\tau}
\def\u{\upsilon}
\def\w{\omega}
\def\x{\xi}
\def\y{\eta}
\def\z{\zeta}
\def\D{\Delta}
\def\G{\Gamma}
\def\H{\Theta}
\def\L{\Lambda}
\def\F{\Phi}
\def\P{\Psi}
\def\S{\Sigma}
\def\me{\mathrm e}
\def\ol{\overline}
\def\tl{\tilde}
\def\*{\dagger}
\def\red#1{\textcolor{red}{#1}}
\def\WY#1{\textcolor{blue}{#1}}
\def\ft#1{\textcolor{red}{#1}}
\def\FT#1{\textcolor{red}{[{\bf FT:} #1}]}


\preprint{TU-1155}

\title{
Power spectrum of domain-wall network 
and its implications for isotropic and anisotropic cosmic birefringence\\
}

\author{
Naoya Kitajima
}
\affiliation{Frontier Research Institute for Interdisciplinary Sciences, Tohoku University, Sendai, Miyagi 980-8578, Japan}
\affiliation{Department of Physics, Tohoku University, 
Sendai, Miyagi 980-8578, Japan} 

\author{
Fumiaki Kozai
}

\affiliation{Department of Physics, Tohoku University, 
Sendai, Miyagi 980-8578, Japan} 

\author{
Fuminobu Takahashi
}
\affiliation{Department of Physics, Tohoku University, 
Sendai, Miyagi 980-8578, Japan}

\author{
Wen Yin
}
\affiliation{Department of Physics, Tohoku University, 
Sendai, Miyagi 980-8578, Japan}

\begin{abstract}
Recently, based on a novel analysis of the Planck satellite data, a hint of a uniform rotation of the polarization of cosmic microwave background photons, called isotropic cosmic birefringence, has been reported. 
The suggested rotation angle of polarization of about $0.2-0.4$ degrees strongly suggests that it is determined by the fine structure constant, which can be naturally explained over a very wide parameter range by the domain walls of axion-like particles. Interestingly, the axion-like particle domain walls predict not only isotropic cosmic birefringence but also anisotropic one that reflects the spatial distribution of the axion-like particle field on the last scattering surface. 
In this {\it Letter}, we perform lattice simulations of the formation and evolution of domain walls in the expanding universe and obtain for the first time the two-point correlation function and power spectrum of the scalar field that constitutes the domain walls. 
We find that while the power spectrum is generally consistent with analytical predictions based on random wall distributions, there is a predominant excess on the scale corresponding to the Hubble radius. Applying our results to the anisotropic cosmic birefringence, we predict the power spectrum of the rotation angles induced by the axion-like particle domain walls and show that it is within the reach of future observations of the cosmic microwave background.
\end{abstract}

\maketitle
\flushbottom

\vspace{1cm}

{\bf Introduction.--}
Topological defects often appear in a variety of new theories beyond the Standard Model (SM). Defects stabilized by topological charges, once created, can alter cosmological evolution, and have various observational impacts. 

Domain walls (DWs)
are produced associated with the spontaneous breaking of discrete symmetry. After the production, they evolve according to the so-called scaling solution, with an average of $\O(1)$ DWs in the Hubble horizon~\cite{Press:1989yh,Garagounis:2002kt,Oliveira:2004he,Avelino:2005kn,Leite:2011sc,Leite:2012vn,Martins:2016ois}. The energy density of such DWs decreases more slowly than non-relativistic matter or radiation, so its fraction of the energy density of the universe increases with time, leading to the notorious cosmological DW problem~\cite{Zeldovich:1974uw, Vilenkin:1984ib}. Various studies have been conducted on the formation of DW, its scaling solutions~\cite{Press:1989yh,Garagounis:2002kt,Oliveira:2004he,Avelino:2005kn,Leite:2011sc,Leite:2012vn,Martins:2016ois}, and emission of gravitational waves~\cite{Vilenkin:1981zs,Preskill:1991kd,Chang:1998tb,Gleiser:1998na,Dufaux:2007pt,Hiramatsu:2010yz,Kawasaki:2011vv,Hiramatsu:2013qaa,Nakayama:2016gxi,Saikawa:2017hiv, Gelmini:2021yzu}. In particular,  stable DWs must have a sufficiently small tension satisfying $\s\lesssim ({\rm MeV})^3$~\cite{Zeldovich:1974uw,Sousa:2015cqa}. DWs with such a small tension are expected to be composed of very light scalar fields.

Not only the DW energy density, but also the temporal and spatial variations in the field value of the scalar that makes up the DW can be observationally important, especially if the scalar mass is sufficiently light. Those DWs behave as background fields that affect the motion and spin of the microscopic particles coupled to the fields.
Indeed, there are various experiments that search for DWs along this line; in the GNOME experiment, 
they are searching for the effects of axion DWs passing through the Earth via the axion coupling with the spin of the SM fermions~\cite{Pospelov:2012mt, Pustelny:2013rza,Afach:2018eze,Afach:2021pfd}.
The probability of encountering DWs can be enhanced if there is a large deviation from the scaling solution~\cite{Battye:1999eq, Friedland:2002qs,Higaki:2016jjh}.

Another interesting observable that is sensitive to the scalar field value is cosmic birefringence (CB), which is the rotation of the polarization plane of the cosmic microwave background (CMB) photon. Such rotation is induced by the motion of an axion-like particle (ALP) coupled to photons~\cite{Carroll:1989vb,Carroll:1991zs,Harari:1992ea}.  Recently, a novel method for extracting the isotropic CB from the CMB data was developed~\cite{Minami:2020odp}, and the results of the analysis using the Planck satellite data suggested a rotation about $0.2-0.4$ degrees with a significance more than $3 \sigma$~\cite{Minami:2020odp, Clark:2021kze, Diego-Palazuelos:2022dsq}. Intriguingly, it strongly suggests that the rotation angle is mainly determined  by the fine structure constant, and it was shown that the ALP DW can naturally explain it in a very wide parameter range~\cite{Takahashi:2020tqv}.\footnote{See Refs.~\cite{Minami:2020odp,Fujita:2020ecn, Mehta:2021pwf,Choi:2021aze, Nakagawa:2021nme, Gasparotto:2022uqo} for analysis using spatially uniform ALP and Refs.~\cite{Gluscevic:2012me,POLARBEAR:2015ktq,Liu:2016dcg,BICEP2:2017lpa} for the constraints on scale-invariant anisotropic CB.}
The ALP DW not only explains the observed isotropic CB naturally but also predicts a characteristic anisotropic CB determined by the  DW configuration on the last scattering surface (LSS).
Future CMB observations are expected to improve the limits on the anisotropic CB by several orders of magnitude~\cite{Pogosian:2019jbt}, so it is very important to precisely determine the form of the anisotropic CB predicted by the ALP DW.

In this {\it Letter}, we determine the two-point function and power spectrum of the scalar field that makes up the DWs, based on
2D lattice simulations of the DW network. 
We use a renormalizable scalar model to follow the evolution of the DWs from the formation to the scaling regime. 
As we will see, numerical results show overall agreement with analytical estimates based on random wall distributions, but we find that there is a pronounced excess in the scale corresponding to the Hubble radius. 
Numerical simulations of the DW evolution in 2D have the advantage of generating large statistics, and the energy density and velocity of DWs in the scaling regime
have been found to be very similar between the 2D and 3D cases~\cite{Kawano:1989mw, Avelino:2005pe, Avelino:2005kn, Leite:2011sc}. Therefore, the power spectrum obtained in this {\it Letter} is expected to be a good approximation of the power spectrum on the LSS in 3D.
By applying this result to the ALP DW explanation of the isotropic CB, we find that the scenario can be fully tested by the future observations of the anisotropic CB.  While we focus on the anisotropic CB generated at the LSS in the main text,  we will comment on the possibility of generating the anisotropic CB from reionization photons in the supplemental material. There we comment on 
oscillons/I-balls~\cite{Bogolyubsky:1976sc,Gleiser:1993pt,Copeland:1995fq,Kasuya:2002zs} found to be formed associated with DWs in our numerical simulations, and discuss various applications of our analysis.

{\bf Setup.--}
As an example of the model leading to the DW formation, let us consider a $\phi^4$ potential with $Z_2$ symmetry, 
\begin{equation}
V[\phi] = \frac{\l}{4} (\f^2-v^2)^2
\end{equation}
with $\l > 0$. There are two degenerate minima 
 at
$
\phi = \pm v,
$
where the scalar mass is given by $m_\phi = \sqrt{2 \lambda} v$.
Clearly, there is a DW solution that connects the two vacua.
The width and tension of the DW are approximately given by $\delta \sim m_\phi^{-1}$ and 
$\sigma \sim m_\phi v^2$, respectively, but
the precise values of these parameters are not relevant for our purpose since we are interested in the statistical
properties of the DW configuration at scales much larger than the width.
In fact, the behavior of DW at such macroscopic scales is not sensitive to those parameters, and our result is applicable to a more generic potential such as the sine-Gordon potential in the case of the axion, which we will consider later.

The equation of motion in the flat Friedmann-Lema\^itre-Robertson-Walker (FLRW) universe,  is given by
\begin{equation}
\ddot{\phi} + 3H \dot{\phi} - a^{-2}\vec{\partial}^2 \phi
+V'[\phi]
= 0,
\end{equation}
where the dot and prime represent the derivative with respect to the cosmic time $t$ and $\phi$, respectively,  $H = \dot{a}/a$ is the Hubble parameter, $a$  the scale factor, and $\vec{\partial}$  the spatial derivative with respect to the comoving coordinate $\vec{x}$.
Here we neglect the gravitational self-interaction of DWs, since 
the DW energy density must be always subdominant to evade the cosmological DW problem.

{\bf A model based on random DW distributions.--} 
For applications to anisotropic CB, we need the two-point correlation function and power spectrum of the scalar field that makes up the DW. It is known that DWs quickly follow the scaling law after their formation.
Here we present analytical results based on a simple model in which infinitely thin DWs obeying the scaling law are assumed to have a uniform probability distribution 
on an arbitrary 1D line in space. The average distance between adjacent walls is given by $(\kappa H)^{-1}$ with $\kappa$ being a
constant of ${\cal O}(1)$~\cite{Takahashi:2020tqv}.
In this random model, the probability of being in the same vacuum at two spatially distant points $\vec{x}_1$  and $\vec{x}_2$ at the same time is given by
$
  \frac{1}{2}\(1+e^{-2\k a H  \ab{\vec{x}_1-\vec{x}_2}}\).
$
This leads to the two-point correlation function 
\begin{equation}
\label{2P}
\vev{\f[\vec x_1,t] \f[\vec x_2,t]}= v^2 e^{-2 \k a H  \ab{\vec{x}_1-\vec{x}_2}}.
\end{equation}
{This result should hold in any spatial dimensions, but now we limit ourselves
to the 2D case.}
By Fourier transforming this two-point function, we obtain the 2D power spectrum as 
\begin{equation}
{ P}[k,t]\equiv \int{d^2 x e^{i \vec{k} \cdot \vec{x}}} \vev{\f[\vec{x},t] \f[0,t] }= \frac{4\pi \k a H v^2}{(4 \k^2 a^2 H^2 +k^2)^{\frac{3}{2}}}. \label{Prm}
\end{equation}
It is convenient to define ${\cal P}[k,t] \equiv k^2 P[k,t]$ which has a maximum at
$k/k_H = \sqrt{2}\kappa/\pi$, where $k_H \equiv 2 \pi a H$ is the wavenumber corresponding to the Hubble horizon.
We
note that the random model is based on the 1D DW distribution and does not reflect information on the global shape of the spatially extended DW, which will be important later to interpret the numerical results. 
The reason why we consider 2D is that CB is sensitive to the power spectrum on the LSS, and we can increase statistics in the 2D case.

{\bf Numerical results.--}
We consider the radiation dominated universe, and
 numerically solve for the rescaled scalar field  
$\hat{\f } =\frac{ a}{v}\f$ on the lattice with the spatial coordinates  $\hat{x}_i= \sqrt{\l} v   x_i$ and the conformal time  $\t= (\sqrt{\l} v t)^{1/2}$. 
We set the scale factor by $a(\tau)=\tau/\tau_{\rm ini}$ where $\tau_{\rm ini}$ is the initial conformal time.
We take the periodic boundary condition in two spatial directions, while the system is assumed to be homogeneous along the remaining one.
As the initial condition of the scalar field at $\tau = \tau_{\rm ini}$, we take the random Gaussian noise around the origin with the probability distribution given by
\begin{equation}
f[\hat\phi_{\rm ini}]= \frac{1}{\sqrt{2\pi \vev{\hat\phi_{\rm ini}^2}}} e^{-\frac{\hat\phi_{\rm ini}^2}{2\vev{\hat\phi_{\rm ini}^2}} },
\end{equation}
and a vanishing time derivative.
Note that the Gaussian distribution represents the quantum vacuum fluctuations in the FLRW spacetime~\cite{Polarski:1995jg, Khlebnikov:1996mc} (see c.f. Refs.~\cite{Khlopov:2004sc, Takahashi:2020tqv} for the DW formation via the inflationary fluctuations).

\begin{figure}[!t]
\begin{center}  
   \includegraphics[width=80mm]{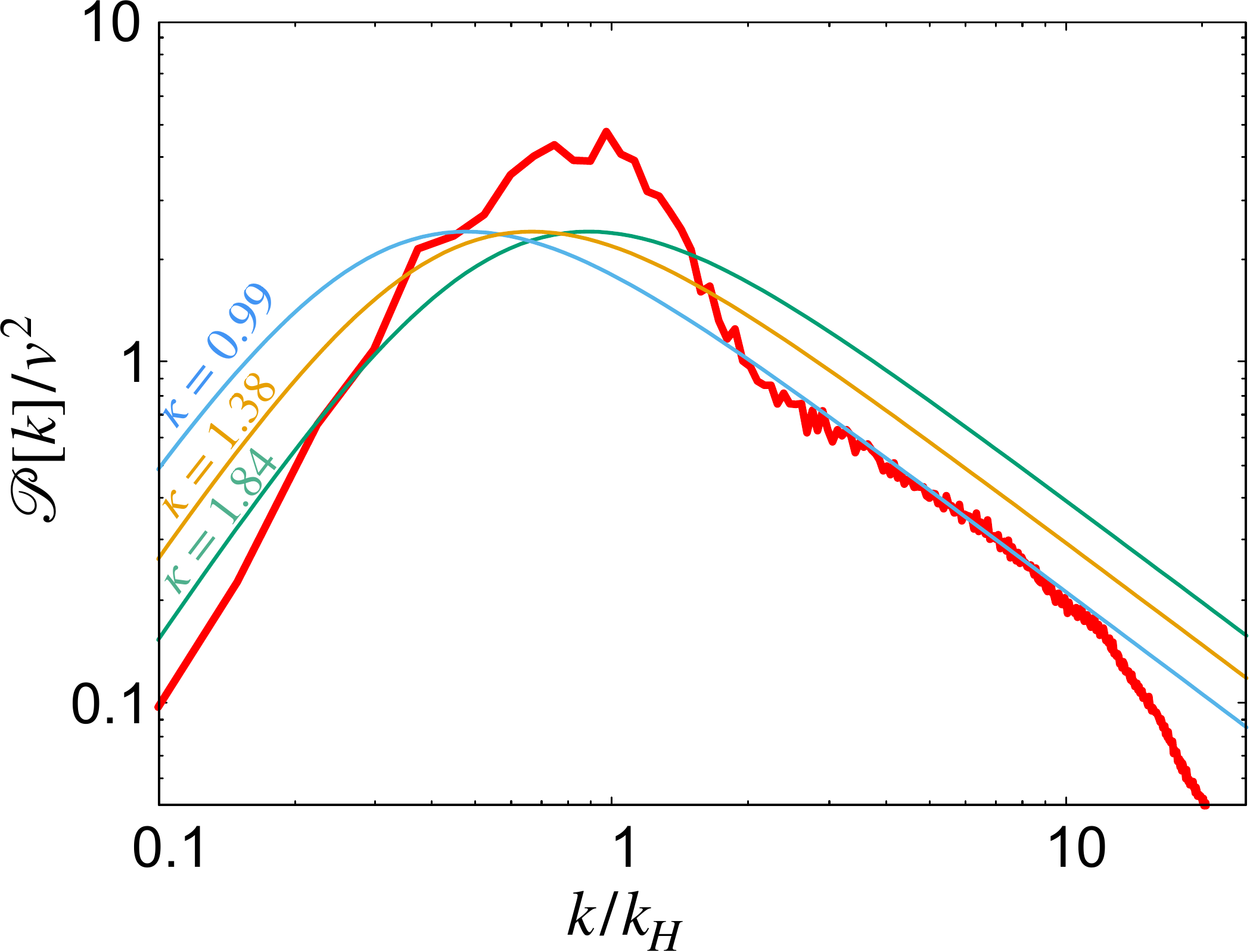}
      \end{center}
\caption{ The numerical result of the power spectrum (thick solid (red) line) at  $\tau=7$. 
The fits of the random model at small, whole, and large scales are shown as 
thin (blue, orange, and green) lines, from left to right. The fits exclude the region of  $k/k_H \gtrsim 11$, where the effects of the DW width and scalar fluctuations are relevant. 
} \label{fig:1}
\end{figure}

In the numerical simulations,
we take the initial conformal time as $\tau_{\rm ini} = 1/2$, 
the comoving box size $L= 100(\sqrt{\lambda} v)^{-1}$,
the lattice spacing $L/N_{\rm lat}$ with $N_{\rm lat}=4096$, and
$\vev{\hat\phi_{\rm ini}^2} = 10^{-12}$.
In Fig. \ref{fig:1}, we show the numerical result of the power spectrum ${\cal P}[k,\tau]/v^2$
obtained by averaging 50 samples evaluated at $\tau=7$.
We have checked that the scaling regime is reached at $\tau\gtrsim 6$ and that the result is essentially the same even if we vary $\vev{\phi_{\rm ini}^2}$ and $\tau_{\rm ini}$ over a certain range. 
Also, our results in the following do not significantly change in the range of $\t=7-10$ as long as we focus on the mode larger than the width of DWs, i.e., $k/a\ll m_\f.$ 

One can see that the numerical result is generally consistent with the random model prediction (\ref{Prm}) at scales $k/k_H\lesssim 11$ where the effect from the wall width can be neglected. 
Fitting the random model prediction to the numerical results for 
$0.1 \lesssim k/k_H \lesssim 11$ yields
\begin{align} \kappa \;{\simeq 1.38}
\end{align}
which is shown as the thin (orange) line in the figure. 
In detail, the best-fit random model prediction is slightly larger than the numerical result at scales larger and smaller than the horizon, while around the horizon scale $k\sim k_H$ there is a clear excess in the numerical results.
{For reference}, fitting the prediction of the random model to scales larger and smaller than the horizon, respectively, yields 
\begin{align}
\kappa \simeq
\left\{\begin{array}{ll}
	\displaystyle{ 1.84~~~~~~~~\text{for}~~0.1  \lesssim k/k_H \lesssim 0.3} &\\
	\displaystyle{ 0.99~~~~~~~~\text{for}~~2 \lesssim  k/k_H\lesssim 11  }
	\label{evolution}
	\end{array}
	\right..
\end{align}
They are shown as thin blue and green lines in the figure.

The scale dependence of $\kappa$ can be attributed to the fact that the characteristic spatial curvature of DW is given by the Hubble horizon. 
{Consider a 1D straight line and count DWs the line passes through. At scales below the horizon, the curvature of DWs is negligible, but at scales around or above the horizon, 
 the straight line will penetrate more DW because the DW is curved.}
 This explains the excess at $k \sim k_H$ and also explains that the superhorizon mode has a larger $\kappa$ than the subhorizon mode.

 As mentioned earlier, at $k\gg 10k_H$, the numerical results fall well below the random model predictions, but this is due to the effect of the finite thickness of the DW. The random model also neglects fluctuations in the scalar field around each vacuum, and in fact, we have confirmed that this effect is sufficiently small.

{\bf 
Cosmic birefringence induced by DWs
--}

Now let us consider that $\f$ is an ALP coupled to a pair of photons via
\begin{equation}
\label{eq:int}
{\cal L}  = -c_\g \frac{\a}{4 \pi} \frac{\phi}{f_\f} F_{\mu \nu} \tilde F^{\mu \nu},
\end{equation}
where $f_\f$ is the decay constant, $c_\g$ a dimensionless model-dependent constant,\footnote{{It is given by $c_\gamma= \sum_{i=1}^{N_\psi} {q}_{{\rm PQ},i} q_{{\rm EM},i}^2/2$ for chiral fermions $\psi_i$ ($i=1\cdots N_{\psi}$) with PQ charge $q_{{\rm PQ},i}$ and electromagnetic charge $q_{{\rm EM},i
}.$} }  $\a$ the fine structure constant, and $F_{\mu \nu}$ and  $\tl{F}_{\mu \nu}$ the field strength and its dual,
 respectively. 
Since the ALP has a periodic potential with periodicity $2\pi f_\phi$, there exists a DW solution connecting two adjacent vacua. If $\f$ has large inflationary fluctuations, multiple vacua and DWs separating them may appear first, but they will likely end up with two vacua and the corresponding DW solution in the course of the DW evolution~\cite{Takahashi:2020tqv}. (See the supplemental material for other possible DW formations).
As long as we consider the DWs separating the two minima and their distribution at macroscopic scales $k/a \ll m_\f$, the previous numerical result
can be applied to this case with the replacement of $v=\pi f_\f$.

The ALP DWs, if exist at the 
recombination epoch, affect the polarization pattern of the CMB photons and induce both isotropic and anisotropic CB~\cite{Takahashi:2020tqv}. The rotation angle of the CMB polarization only depends on the difference in the ALP field value between the LSS and the Earth,\footnote{In contrast, the rotation angle does depend on the path in the presence of cosmic strings with/without DWs~\cite{Agrawal:2019lkr, Jain:2021shf, Yin:2021kmx}. This is because the ALP field is ill-defined and singular at the core of the strings. In this case, the induced CB is predominantly anisotropic, determined by the string distribution after the recombination era,
and its magnitude is related to $\alpha$~\cite{Agrawal:2019lkr}.
}
 and it is given by~\cite{Carroll:1989vb,Carroll:1991zs,Harari:1992ea},
 \begin{equation}
 \laq{polar}
 \D\F(\Omega) \simeq 
  0.42 {\rm~ deg } \times c_\g \( \frac{\f_{\rm today}-\f_{\rm LSS}(\Omega) }{2\pi f_\phi}\),
 \end{equation}
where $\f_{\rm today}$ and $\f_{\rm LSS}(\Omega)$  are the ALP field value at the Earth today, and at the LSS, respectively,
and $\Omega$ denotes the angular direction specified by the polar coordinates $(\theta,\varphi)$. In the following we set $\f_{\rm today}-\f_{\rm LSS}(\Omega) \geq 0$ without loss of generality. Note that it is  $\alpha$ that determines the numerical factor, since $1/137 {\rm\, rad} \simeq 0.42 {\rm\, deg}$.

The two vacua are {equally populated} at the LSS, but the observer in the solar system
is in one of them. Thus, the symmetry that exchanges these two vacua appears to be spontaneously broken by the observer.
Then, the rotation angle $\Delta \Phi(\Omega)$ is 
either $0$ or $0.42 c_\gamma$ deg, depending on which vacuum is realized at the LSS in the corresponding angular direction.
For DWs following the scaling solution, the pattern of the CB is characterized by $2^N$, i.e. $N$-bit, of information with 
$N=\O(10^{3-4})$
being equal to the number of domains at the LSS, and is called kilobyte CB (KBCB)~\cite{Takahashi:2020tqv}.

By the fact that the observer  is in either vacuum, the KBCB predicts the isotropic rotation angle $\beta$ as
\begin{equation}
\beta \simeq  0.21 c_\g  {\rm~ deg },
\end{equation}
independent of the ALP mass and decay constant.
Interestingly, for $c_\g=\O(1)$, the prediction is consistent with the observed value~\cite{Diego-Palazuelos:2022dsq} (see also \cite{Minami:2020odp,Eskilt:2022wav}),
\begin{equation} \beta=0.36\pm 0.11 {\rm\, deg},
\end{equation}
for a wide range of the ALP mass and decay constant.

In addition, the KBCB predicts characteristic anisotropic CB that reflects the DW configuration on the LSS, and its magnitude is naturally correlated with the isotropic one. To evaluate the anisotropic CB induced by DWs, we use the flat-sky approximation to translate our 2D lattice result given above.
This should work since various properties of the DW network such as the exponent of the energy density and the velocity as well as the correlation length are similar in the 2D and 3D cases, and their difference is within ${\cal O}(10)\%$~\cite{Kawano:1989mw, Avelino:2005pe, Avelino:2005kn, Leite:2011sc}. 
Based on the random model in 3D, 
the two point function for the anisotropic CB  is given as \cite{Takahashi:2020tqv}
 \begin{equation} 
 \label{aCB}
 \vev{ \Delta \tl{\F}(\theta, 0) \Delta \tl{\F}(0,0)} =   \beta^2  e^{-2 \k a_{\rm L} H_{\rm L} d_{\rm L} \sqrt{2(1-\cos\theta)}}
  \end{equation} 
where $\D \tl\F\equiv \D\F -\b$ is the fluctuation of the rotation angle,
$a_L$ and $H_L$ are the scale factor and the Hubble parameter at the LSS, and $d_L$ is the comoving distance  to the LSS. We define the angular power spectrum as
\begin{equation}
\label{Cell}
C_\ell^\Phi = 2\pi \int_{-1}^{1} d \cos \theta \vev{ \Delta \tl{\F}(\theta, 0) \Delta \tl{\F}(0,0)} P_\ell(\cos \theta)
\end{equation}
with $P_\ell$ being the Legendre polynomial. For $|\theta| \ll 1$, we can use the flat-sky approximation where the Legendre polynomial is approximated by the Bessel function of the first kind of order $0$,  and the precision is sub-percent level for $\ell \gtrsim 10$~\cite{Bernardeau:2010ac}.
Comparing Eqs.~(\ref{aCB}) and (\ref{Cell}) with Eqs.~(\ref{2P}) and (\ref{Prm}) under the flat-sky approximation, 
we obtain the correspondence,
\begin{equation}
\laq{map} 
\sqrt{\ell (\ell+1)}/d_L \leftrightarrow k,~~~~ \ell (\ell+1) C_\ell^\F/\beta^2 \leftrightarrow  {\cal P}[k]/v^2.
\end{equation}

Other possible errors in this translation are as follows. The numerical simulations described above assumed radiation dominance. In the actual universe, there is a short matter-dominant period before recombination. We have also performed simulations for the matter-dominated universe and found that the power spectrum does not change much near the Hubble scale. In addition, DWs typically have a velocity of about half the speed of light and also generally form an $O(1)$ angle with the LSS. Since the thickness of the LSS is about $0.1 /H_{\rm L}$, their effects will be significant for  $\ell \gtrsim 1000$, but subdominant for $\ell \lesssim 1000$. In summary, we expect that the total error in our estimate using the numerical simulations is of ${\cal O}(10)$\% for the scales of our interest.

We show the predicted anisotropic CB 
as the solid (red) line in Fig.~\ref{fig:2} where we use $\beta=0.36$ deg as suggested by the recent analysis (the predicted spectrum scales with $\beta^2$). The upper and lower thin (red) lines are the prediction multiplied, respectively, by 2 and 1/2 to show the expected uncertainty. {The cosmic variance is given by $\sqrt{2/(2\ell+1)}C_\ell^\F$~\cite{Takahashi:2020tqv}, but is not shown in the figure because it is negligibly small.}
We also display the future reaches of LiteBIRD~\cite{Matsumura:2016sri}, Simons Observatory~\cite{Ade:2018sbj}, CMB-S4 like experiment~\cite{Abazajian:2016yjj}, and PICO~\cite{Hanany:2019lle} by the blue lines. Here we adopt the statistical uncertainty given in Ref.~\cite{Pogosian:2019jbt},\footnote{
The uncertainty was estimated assuming vanishing isotropic CB.
The nonzero isotropic CB may improve the sensitivity of future CMB observations on the anisotropic CB. 
We thank Toshiya Namikawa for the comment on this issue.} where the assumed beam systematics are such that the scientific targets are achieved in those experiments.
Even if $\b$ is near the lower end of the $1\s$ range suggested by the CMB observations, the predicted anisotropic CB in our scenario can be tested by the future CMB-S4 and PICO experiments.

\begin{figure}[!t]
\begin{center}  
   \includegraphics[width=75mm]{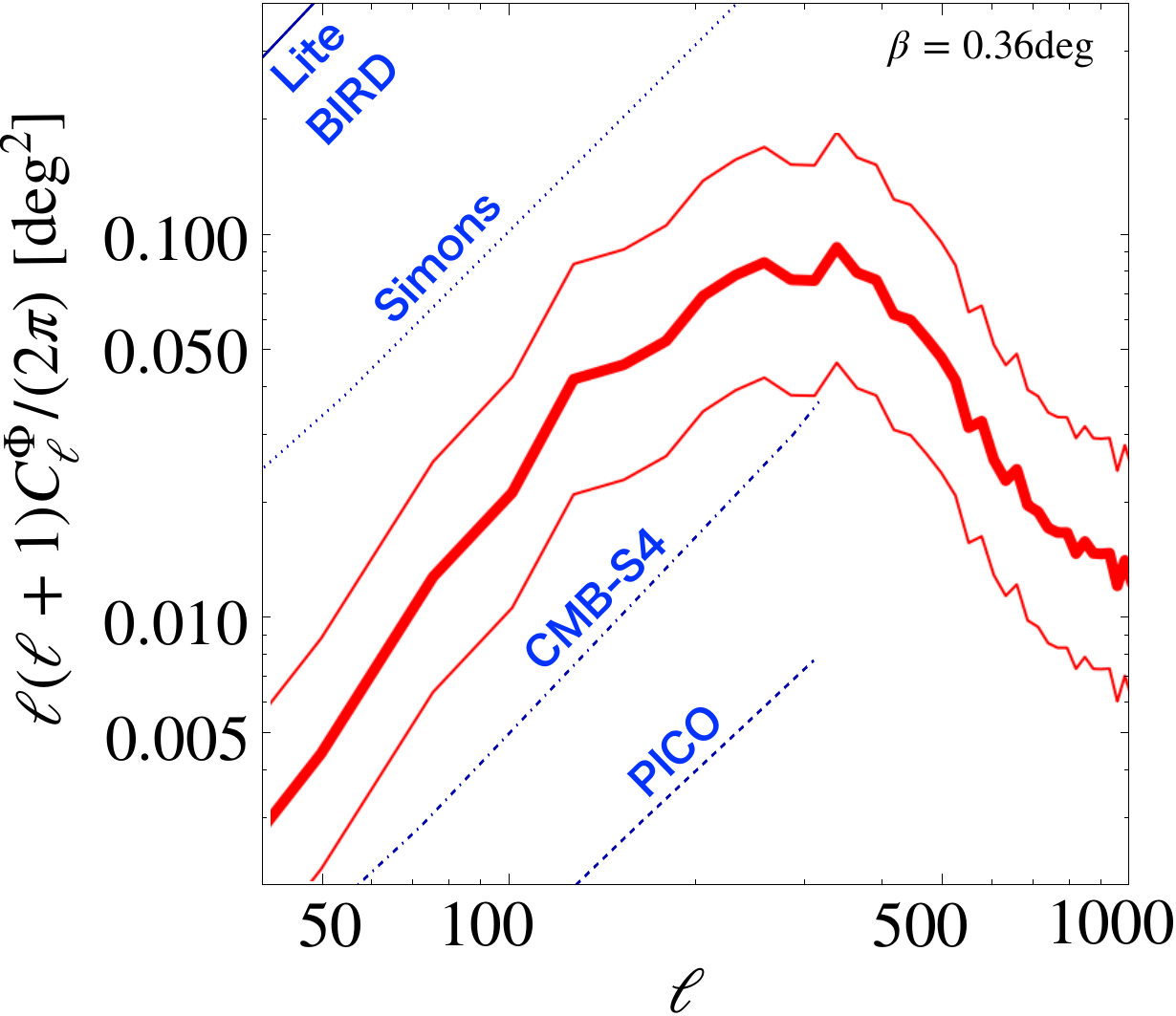}
      \end{center}
\caption{Anisotropic CB caused by ALP DWs without strings [red solid line], predicted based on 2D numerical lattice calculations. 
Here we fix $\beta=0.36\,$deg.
To take account of the $\O(10)\%$ uncertainty discussed in the main text we also show the results multiplied by $1/2,2 .$ 
The future reaches of LiteBIRD~\cite{Matsumura:2016sri}, Simons Observatory~\cite{Ade:2018sbj}, CMB-S4 like experiment~\cite{Abazajian:2016yjj}, and PICO~\cite{Hanany:2019lle} are shown in blue lines from top to bottom. They are adopted from \cite{Pogosian:2019jbt}. 
} \label{fig:2}
\end{figure}

{\bf 
Conclusion
--}
We have evaluated the power spectrum of the anisotropic CB induced by the ALP DWs based on the numerical simulations, and have shown that it is within the sensitivity of future CMB observations.  It would be important to evaluate other correlation functions. In particular, some of the cross bispectrum with CMB anisotropies have a large signal-to-noise ratio~\cite{Greco:2022ufo}, which is worth studying in our case.\footnote{We thank Alessandro Greco for bringing the reference to our attention.}

\section*{Acknowledgments}
We are grateful to Alessandro Greco for informing us of the importance of cross bispectrum with CMB fluctuations.
We thank Eichiro Komatsu and Toshiya Namikawa for fruitful discussions during the symposium of ``What is dark matter? - Comprehensive study of the huge discovery space in dark matter".
This work is supported by JSPS Core-to-Core Program (grant number: JPJSCCA20200002) (F.T.),  JSPS KAKENHI Grant Numbers 17H02878 (F.T.), 19H01894 (N.K.), 20H01894 (F.T. and  N.K.), 20H05851 (F.T.,  N.K. and W.Y.), 21H01078 (N.K.), 21K20364 (W.Y.),  22K14029 (W.Y.), and 22H01215 (W.Y.).

\clearpage
\appendix
\section{Supplemental material}
\subsection{Evolution of DWs and oscillons}
\paragraph{Time-evolution and scaling behavior}
In this supplemental material, we show some of the relevant dynamics of DWs obtained from lattice simulations. 
The scalar field configurations at $\tau =0.5,7$, and  $10$ on the comoving 2D lattice are shown in Fig.~\ref{fig:snap}.  
\begin{figure}[!t]
\begin{center}  
   \includegraphics[width=90mm]{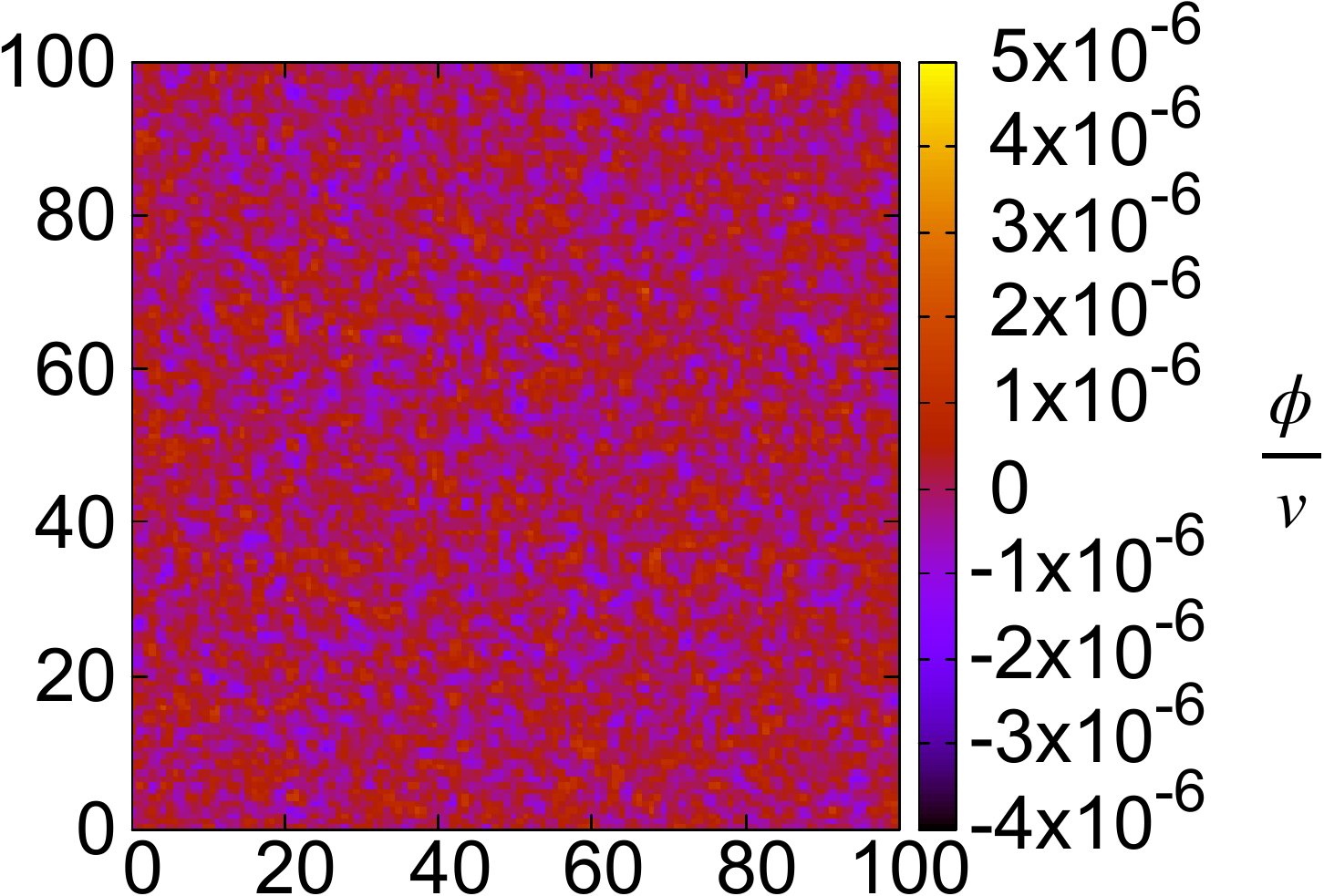}
      \includegraphics[width=80mm]{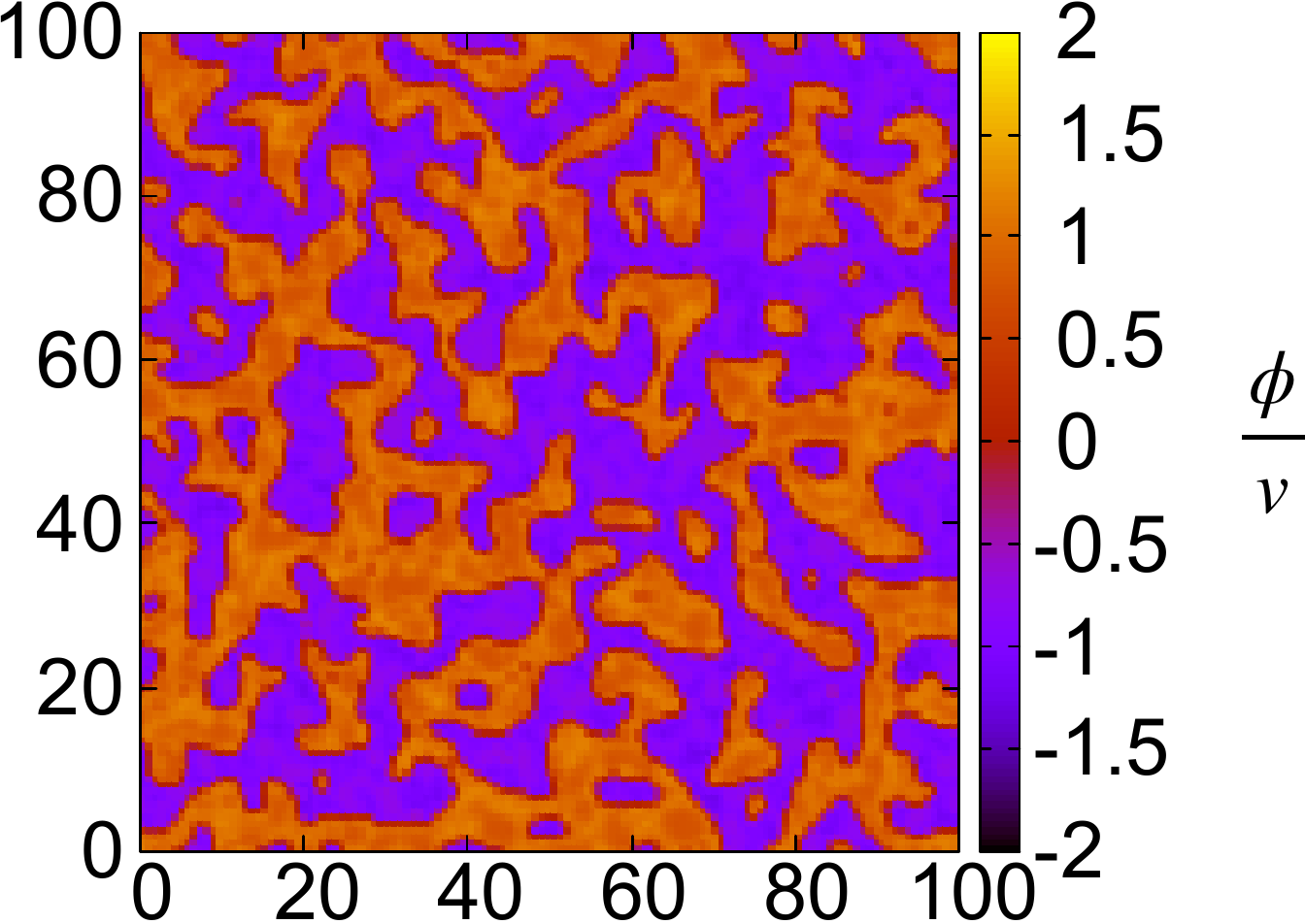}
   \includegraphics[width=80mm]{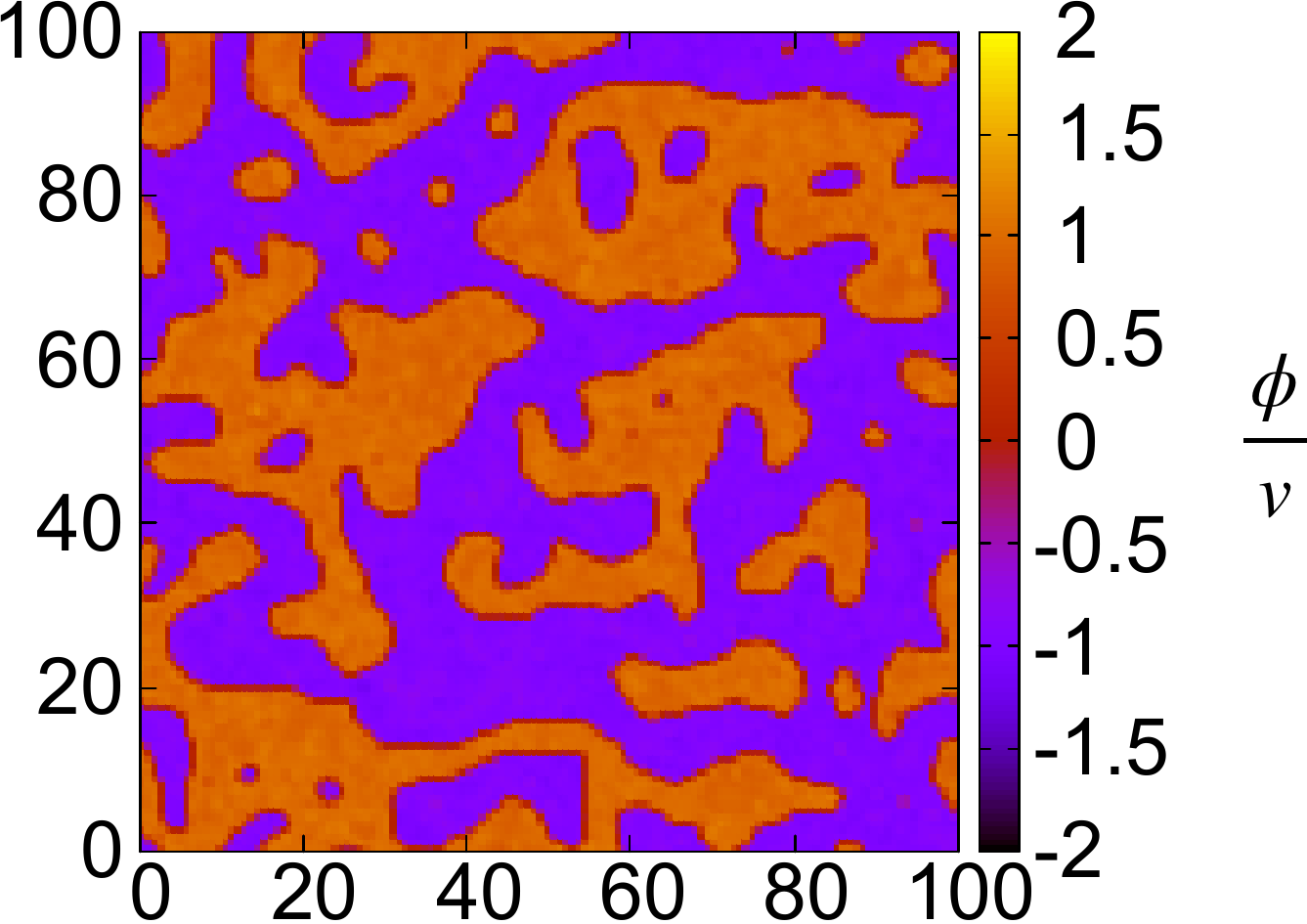}
      \end{center}
\caption{The spatial field configurations of $\f/v$ at $\tau=0.5,7$, and $10$ is shown   from top to bottom. The choice of the parameters is same as in the main text.
} \label{fig:snap}
\end{figure}

As an algorithm for identifying DWs on the lattice, we searched for places where the sign of the scalar field $\phi$ is inverted between adjacent points. The length of the DWs was then measured by adding up the number of such points. As a result, we found that the length of the DWs becomes proportional to the number of the Hubble horizons in the box, i.e. the DWs obey the scaling law, after $\tau \approx 6$. Therefore, the panels of  $\tau=7$ and $10$ in Fig.~\ref{fig:snap} represent the DW network that obeys the scaling law.

\paragraph{Scalar waves and oscillons from DW formation}
In a DW network that evolves according to the scaling law, the DWs are constantly annihilating. During this nonlinear evolution process, the energy stored in the DWs is released in some form. We show in Fig.\ref{fig:osc} a magnified view of the distribution of scalar fields and energy density at $\tau=7$. Comparing the two panels, we can see that there are many point-like lumps of energy, where the field values are near the potential minimum. This means that the energetic objects are not DWs, but excited states around the potential minimum. These point-like objects could be oscillons/I-balls. We can also see scalar waves that may be induced by the annihilation or motion of the DW.

Those energetic objects, together with the DW annihilation, may be emitting gravitational waves characteristic of their respective scales. Observing such gravitational waves may be challenging because there is a strict upper limit on the energy density of DW to avoid the cosmological DW problem in the case of stable DWs. On the other hand, one should keep in mind that these phenomena could be new observational probes of DWs toward the discovery of DW. Also, if the DW is unstable and disappears at a certain point, such objects could remain for a long time and have cosmological consequences.

\begin{figure}[!t]
\begin{center}  
 \includegraphics[width=90mm]{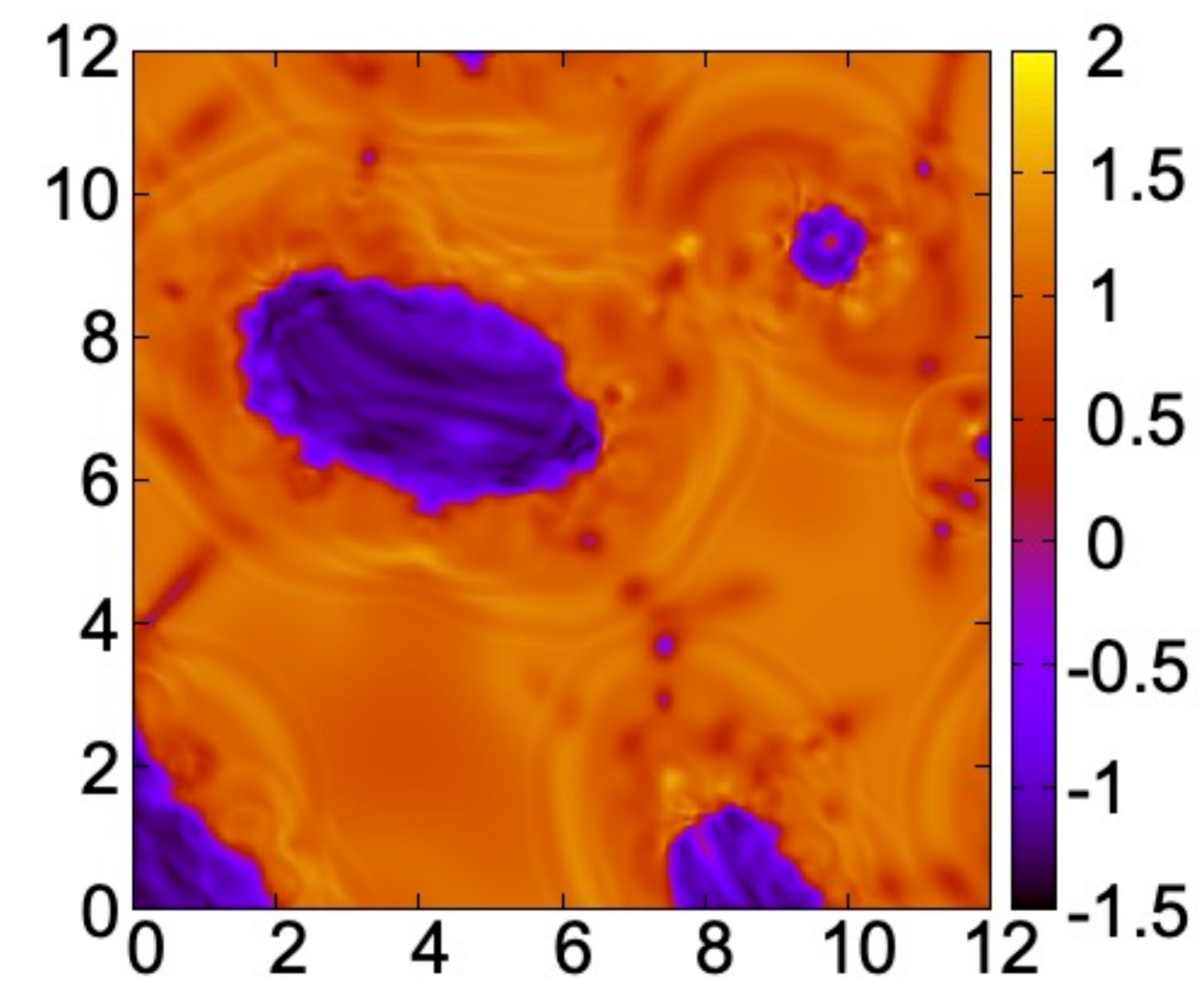}
  \includegraphics[width=90mm]{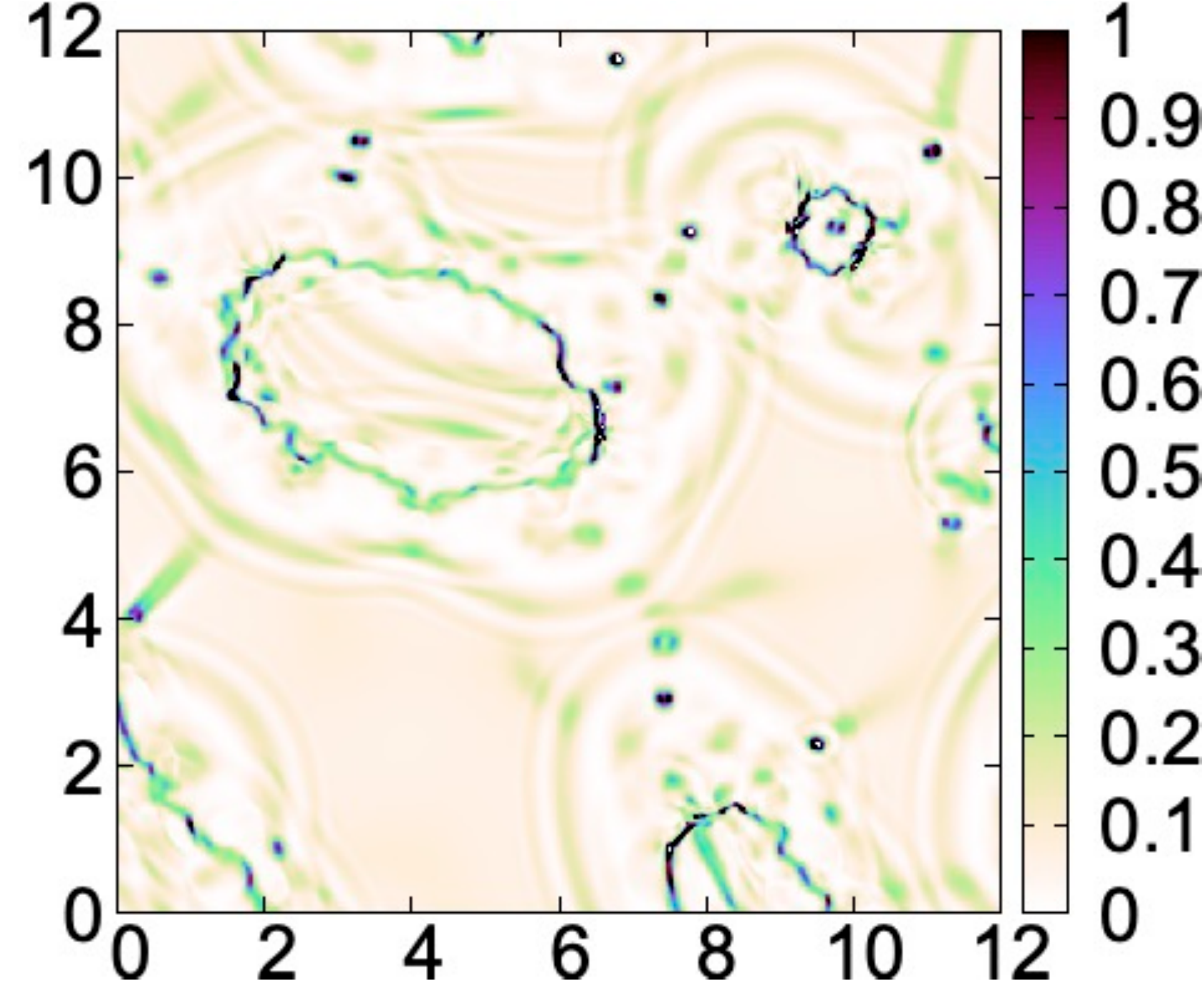}
       \end{center}
 \caption{Enlarged view of the distribution of scalar field values $\phi/v$ (upper panel) and energy density $\rho/m_\phi^2 v^2$(lower panel) at $\tau=7$.
 One can see that there are many point-like objects that appear to be  oscillons/I-balls.
} \label{fig:osc}
\end{figure}

\subsection{Mechanisms of ALP DW formations}

In this supplemental material, we discuss some other possibilities than the inflationary production for the DW formations in the context of CB to strengthen the conclusions in the main part. 

It is possible to shift the axion potential by considering the 
inflaton-axion mixing~\cite{Daido:2017wwb,Takahashi:2019pqf,Takahashi:2019qmh,Nakagawa:2020eeg} (See also \cite{Co:2018mho,Kobayashi:2019eyg, Huang:2020etx}). If the shift is equal or sufficiently close to $\pi$, we can 
effectively flip the sign of the potential. We can set the ALP distribution across the 
potential maximum if it is initially around the potential minimum. This provides a similar initial condition as our numerical simulation. 

In addition, one may also consider the axion (or ALP) roulette~\cite{Daido:2015bva, Daido:2015cba}, where an ALP rotates around the periodic direction via the multi-axion mixings and level-crossing. It was found that the DWs can be formed in this process. This scenario is interesting because if such a fast rotating ALP exists around the electroweak symmetry breaking, and if the photon coupling (partially) originates from W-boson coupling, a sizable chemical potential for the $B+L$ number is obtained, 
$\mu_{\rm B+L}\sim  \frac{\dot{\phi}}{f_\phi}$. 
To have the baryogenesis, this should be around $ 10^{-9}T$ at the electroweak symmetry breaking. 
Conversely, the DW problem may exist in this kind of baryogenesis scenario. Avoiding the problem by a small mass of the ALP may predict the CB. 
A more detailed analysis of this possibility will be discussed elsewhere.

\subsection{CB from photons scattered during reionization}
In the main text, we have focused on CMB photons traveling straight from the LSS.
Recently it was pointed out in Refs. \cite{Sherwin:2021vgb, Nakatsuka:2022epj} that the photons scattered during reionization (at $z \sim 8$ whose Hubble radius corresponds to the scale for the multipole moment $\ell\sim 10$) allow a tomographic approach in the early universe to probe the isotropic CB. We note that the isotropic CB from the reionization does not change in the KBCB scenario. This is because it is the ``spontaneous" $Z_2$ breaking of the vacuum at the solar system that is responsible for the isotropic CB, while $\vev{\f}$ on the reionization surface or LSS  is almost zero. On the other hand, the anisotropic CB from the reionization photons carries important information. This is because the scaling behavior of DWs tells us that there should be a contribution to the power spectrum like Fig. 2 in the main part with $d_L$ replaced with the distance to the reionization surface, which has a peak at $\ell \sim 10$. This does not change much the power spectrum at smaller scales.

Another interesting possibility is that when the scalar mass is in the range of $10^{-32}\EV<m_\f<10^{-28}\EV$, 
the peak is only at the reionization scales since the DW formation is after the recombination. In addition, there is a scale-invariant spectrum from the CMB photons from LSS if the initial $\phi$ distribution before the DW formation has inflationary fluctuations. Therefore, the information on the CMB photons scattered by reionization may be used to shed light on the masses of the scalar fields that make up the DW and the timing of DW formation.

{On the other hand}, we may consider some biased potential or biased initial condition
so that DWs disappear soon after the recombination. In this case, the CMB photons scattered at the LSS have anisotropic CB as in the main text, but the contribution from the photons scattered during reionization is suppressed. 
Thus, we can distinguish various scenarios by measuring the CB of the photons scattered during reionization.

\bibliography{ref}

\end{document}